# Anisotropic magnetocaloric effect in all-ferromagnetic ($La_{0.7}Sr_{0.3}MnO_3$/$SrRuO_3$) superlattices

S. Thota, Q. Zhang, F. Guillou, U. Lüders, N. Barrier, W. Prellier[*] and A. Wahl

Laboratoire CRISMAT, CNRS UMR 6508, ENSICAEN,

6 Boulevard du Maréchal Juin, F-14050 Caen Cedex, France

P. Padhan

Department of Physics, Indian Institute of Technology Madras, Chennai-600036, India

## Abstract

We exploit the magnetic interlayer coupling in $La_{0.7}Sr_{0.3}MnO_3$/$SrRuO_3$ superlattices to realize a crossover between inverse and conventional magnetic entropy changes. Our data reveal a strong anisotropic nature of the magnetocaloric effect due to the magnetic anisotropy of the superlattice. Therefore, artificial superlattices built from ferromagnetic materials that can be used to alter the magnetic structure as well as the magnetic anisotropy, could also be utilized for tuning the magnetocaloric properties, which may open a constructive approach for magnetic refrigeration applications.

* Author to whom any correspondence should be addressed

In the recent years, magnetocaloric effect (MCE) in mixed-valency manganese-oxides[1] have been extensively studied owing to the fact that some of them exhibit fairly large entropy change ($\Delta S_M$ values) comparable with Gd- based alloys. Of particular interest are "AMnO$_3$" type structures where "A" is trivalent rare-earth ion mixed with divalent alkaline-earth element (e.g. La$_{1-x}$Sr$_x$MnO$_3$) because they exhibit a rich variety of magnetic and electronic properties.[2] However, the main drawback in these systems is large heat capacities leading to small temperature change ( $\left| \Delta T \right|$ ~2K for a field change, $\Delta H$ of 2T).[3,4] On the other hand, a wide range of transition temperatures (100 K $\leq T_C \leq$ 375 K) together with large MCE values (associated with the sharp rise of the magnetization) for small field changes are the positive features which makes them useful in magnetic refrigeration.[5] Most of today's research activity in this field is limited to bulk materials. Little literature is available on the MCE properties of thin films[6-8], although principally they will be easier to integrate into electronic structures for applications. Nevertheless, the concept of using magnetic superlattices for magnetic cooling has been recently illustrated by Mukherjee et al. in Co/Cr superlattices.[7] They suggested that the interlayer antiferromagnetic coupling between two ferromagnetic constituents-(Co) separated by a non magnetic layer (Cr) in their case-can be used to tune the MCE properties, and found that Co/Cr superlattices exhibit a maximum entropy change of >-0.4 J kg$^{-1}$ K$^{-1}$, at 330 K. Apart from interlayer coupling, superlattices can show strong magnetic interface coupling when two ferromagnetic materials are in close contact with each other resulting in antiferromagnetic coupling[9,10] which would lead to the growth of magnetic structures comparable to those proposed by Mukherjee et al.[7] In this direction, we have explored the magnetocaloric properties in all-ferromagnetic (FM) La$_{0.7}$Sr$_{0.3}$MnO$_3$/SrRuO$_3$ (LSMO/SRO) artificial superlattices. Our results indicate (i) the presence of interfacial antiferromagnetic (AFM) coupling between the FM materials $SrRuO_3$ and $(La,Sr)MnO_3$,[11] and (ii) an inverse (negative) and conventional (positive) anisotropic-MCE, below and above the temperature at which this AFM coupling appears.

A multitarget pulsed laser deposition system operating at 248 nm wavelength has been used. The bottom layer LSMO (20 unit cells) is directly grown on the [001]-oriented $SrTiO_3$ substrate followed by 'n' unit cells of SRO layer (in the present case 'n' = 6). The above bilayer is repeated 15 times and finally covered with 20 unit cells of LSMO. The details of preparation conditions and their structural analysis have been reported elsewhere.[12] A superconducting quantum interference device based magnetometer (Quantum Design MPMS) with magnetic field (H) up to 50 kOe and temperature (T) 5-340 K has been used to perform the magnetization (M) measurements.

Fig. 1 shows the M(T) plots measured in the presence of a 50 Oe magnetic field applied along the out-of-plane (H//[001]) and in-plane (H//[100]) directions under zero-field-cooling (ZFC) and 50 Oe field-cooling (FC) conditions. A good estimation of the transition temperatures can be drawn from the $[\partial M/\partial T]$ versus T plot shown in the inset of Fig. 1(a). For both in-plane and out-of-plane configurations, the first transition at 325 K and the corresponding magnetic moment are coherent with the Curie temperature ($T_C^{LSMO}$) and bulk magnetic moment of $La_{0.7}Sr_{0.3}MnO_3$, respectively.[13] The $T_C$ of $SrRuO_3$ ($T_C^{SRO}$) is not clearly visible in Fig. 1(a) and 1(b) as expected at 145 K where an enhancement of magnetization ($\Delta M \approx 50$ emu cm$^{-3}$) is observed with decreasing the temperature. This $\Delta M$ value is lower than the bulk due to the small volume fraction of $SrRuO_3$. In the case of in-plane configuration the enhancement of the magnetization value within the temperature range $T^* < T < T_C^{SRO}$ is consistent with the onset of ferromagnetic coupling arising from the ordering of the $SrRuO_3$ magnetic moment as emphasized in Ref. 14 (where $T^*$ is the onset of the decrease in the magnetization in the ZFC mode associated with the occurrence of the interfacial magnetic coupling). Further, the in-plane magnetization curves exhibit higher values than that measured in the out-of-plane (~12 times larger at 5K) configuration. This indicates that the easy axis orientation in these superlattices is along (100) direction in contrast to thicker $SrRuO_3$ coherently grown on $SrTiO_3$ substrates which exhibit typically an easy axis normal to the surface due to the strong magnetocrystalline anisotropy of this material.[15]

Below 100 K the decrease in the in-plane $M_{FC}$ and splitting between the $M_{FC}$ and $M_{ZFC}$ indicates the presence of a third magnetic transition with an antiferromagnetic character (indicated by $T^*$ in the inset of Fig. 1(b)). This feature suggests that there is a coupling between $SrRuO_3/La_{0.7}Sr_{0.3}MnO_3$ yielding to a partial compensation of the overall magnetic moment of the superlattice for both in-plane and out-of-plane configuration.[6,12] The value of this compensation, along the easy axis, of around 100 emu $cm^{-3}$ is consistent with a scenario, where the $SrRuO_3$ layer couples antiferromagnetically to the $La_{0.7}Sr_{0.3}MnO_3$.[12] In general, field cooling from a paramagnetic (disordered) state to a lower temperature gives preferential orientation to the spins, whereas in zero field cooling no preferred orientation is induced. Thus, cooling the sample in the presence of a small magnetic field prevents the antiferromagnetic interlayer coupling between the $La_{0.7}Sr_{0.3}MnO_3$ and $SrRuO_3$.

Fig. 2(a) shows the magnetic hysteresis loops recorded at 70 K applying the magnetic field along [001] (i.e. out-of-plane) and the [100] (i.e. in-plane) directions after subtracting the substrate diamagnetic contribution (which is estimated from extrapolating the high field data). The values of the remanent magnetization ($M_r$) and coercive fields ($H_C$) under the in-plane direction are 441 emu $cm^{-3}$ and 198 Oe, respectively; where as for the out-of-plane direction $M_r$ and $H_C$ being 48 emu $cm^{-3}$ and 400 Oe, respectively (clearly shown in the inset of Fig. 2(a)). The above observations confirm that the [100]-direction is an easy axis, the [001] is a hard axis. As stated before, the magnetic moments of the $SrRuO_3$ are thus forced into the sample plane. In these superlattices, the small thickness of the layers (6 unit cells) and strain at the interfaces could significantly modify the magnetic anisotropy in $SrRuO_3$ films[15].

Fig. 2(b) shows the in-plane magnetization at low magnetic fields at different temperatures. Prior to each measurement, the temperature was raised above the $T_C^{LSMO}$, in order to ensure a perfect demagnetization. The value of in-plane magnetization per unit volume in each of the isotherms increases gradually until a transition magnetic field ($H_T$), where it rises rapidly with the applied magnetic field, and then saturates (See Fig. 2(b)). The transition field ($H_T$) corresponds to the point of maximum slope in the magnetization isotherm. First the magnitude of the magnetization step ($\Delta M$) has been estimated using the difference between the linear extrapolations of M(H) curve above and below the transition region. The field corresponding to the midpoint of the '$\Delta M$' gives the rough estimation of $H_T$. This can be understood on the basis of the AFM coupling of the SRO/LSMO films. The spontaneous magnetization at H = 0 corresponds to a partially compensated overall moment ($M_{LSMO}$-$M_{SRO}$), and the application of a threshold magnetic field leads to the alignment of the magnetic moment of $SrRuO_3$ in the field direction and therefore, saturates at a value of ($M_{LSMO}$+$M_{SRO}$). Such behavior closely resembles a metamagnetic-like transition, where a jump in the magnetization value ($\delta M$) results from the suppression of AFM-like coupling at the interface. This feature is induced by changing the field strength in the vicinity of this first order transition (either in M versus H or M versus T).[16] The temperature-dependence of '$H_T$' is shown in the inset of Fig. 2(b). With increasing the temperature, '$H_T$' gradually decreases and approaches zero near $T_N$ where AFM coupling is totally suppressed. $H_T$ is a measure of the strength of the antiferromagnetic coupling. The low values of $H_T$ indicate a weak antiferromagnetic interlayer/interfacial coupling at low temperatures. This particular low field behavior was only observed in the virgin M-H curves, and not in the hysteresis loops, due to this extremely small coupling strength. The application of a high magnetic field destroys the interface coupling, so these hysteresis loops have the same shape as for a typical ferromagnet. The M(H) measurements recorded in the out-of-plane configuration are shown in Fig 2(c). Due to its hard axis nature, a stronger field is necessary to break the interfacial coupling and, to align the magnetic moment of the $SrRuO_3$ in the field direction. This is the reason why one does not observe this feature at low temperature within the investigated field range.

From the isothermal M(H) curves (Figs. 2(b,c)), the change of magnetic entropy, $[-\Delta S_M]$ is estimated at various magnetic field changes ($\Delta H$) using both Maxwell's $[\partial S^{(T,H)}/\partial H]_T = [\partial M^{(T,H)}/\partial T]_H$ and Clausius-Clapeyron (CC) equations $= \Delta S^{(T)} = \delta M \times [\partial T/\partial H_T]^{-1}$.[17,18] In the present case '$\delta M$' occurs under the application of a small magnetic field ($\leq$ 400 Oe) in each isotherm which in turn causes variation in the magnetic entropy. This is an important aspect since such significant variation in the entropy by means of moderate magnetic fields will indeed be more appropriate than bulk for magnetic refrigeration technology.[17,19]

The behavior of $[-\Delta S_M]$ (Figs. 3(a,b)) strongly depends on field orientations, temperature and the magnitude of field change ($\Delta H$). In the in-plane configuration (easy magnetic axis), we observe small negative values in $[-\Delta S_M]$ for a broad range of investigated temperatures for lower applied magnetic fields (at least up to 1T). In this low field regime one can see a crossover of $[-\Delta S_M]$ to positive values for temperatures beyond $T^*$. When the field change is sufficiently high, positive values in $[-\Delta S_M]$ are observed over the whole range of temperatures due to the complete field induced suppression of AFM aligned spins at interface. It is worth pointing out that under high field change the in-plane $[-\Delta S_M]$ values reach 0.85 and 1.3 mJ cm$^{-3}$ K$^{-1}$ for $\Delta H = 1T$ and $2T$, respectively around $T_C$ of SRO. For the bulk SRO, the reported $[-\Delta S_M]$ value around $T_C$ is $\sim 3.83$ mJ cm$^{-3}$ K$^{-1}$ for $\Delta H = 1.5$ T[20] in agreement with the reported bulk data 0.87 mJ cm$^{-3}$ K$^{-1}$ (considering the fact that SrRuO$_3$ occupies 23 % of total volume in the superlattices).

These results provide convincing evidence that below 150 K the main contribution to the total entropy change comes from the SrRuO$_3$. Furthermore, the magnetic entropy changes obtained from Maxwell method are in good agreement with that of CC equations around the metamagnetic transition, up to the field change of 0.5 kOe. In the case of out-of-plane configuration, $[-\Delta S_M]$ reaches a maximum value of -0.6 mJ cm$^{-3}$ K$^{-1}$ at T = 100 K, for a field change of $\Delta H = 2.2$ T (Fig. 3(c)). This field configuration is the hard axis direction of magnetization and consequently; the magnetization saturates at much higher fields than in the in-plane direction, making difficult any comparison with $[-\Delta S_M]$ values at low fields. However, above 2T, small $[-\Delta S_M]$ values are observed for the in-plane configuration, whereas negative values in $[-\Delta S_M]$ have been observed in out-of-plane configuration. The magnitude of entropy changes determined under easy-axis orientation below $T_C^{SRO}$ is greater than that of hard-axis orientation whereas an opposite trend is observed beyond $T_C^{SRO}$. All these results reveal strong anisotropic nature of magnetocaloric properties.

In summary, we have demonstrated that La$_{0.7}$Sr$_{0.3}$MnO$_3$/SrRuO$_3$ superlattices exhibit both inverse and conventional magnetocaloric effects associated with the antiferromagnetic and ferromagnetic coupling, respectively. The sign of the magnetic entropy change depends strongly on the temperature. Furthermore, due to very high magnetic anisotropy of the superlattices, entropy change also depends on the magnitude and direction of applied field strength. Thus, these results reveal strong anisotropic nature of magnetocaloric properties. This may introduce a more efficient way of tuning the magnetocaloric properties through a moderate applied magnetic field (≤400 Oe).

This work is done in the frame of the LAFCIS. Support of the CEFIPRA/IFPCAR (3908-1), the C'Nano Nord Ouest (GDR 2975) and STAR (21465YL), ENERMAT (2009-1/086) and SOPRANO (PITN-GA-2008-214040) projects are acknowledged. We also thank J. Lecourt, L. Gouleuf, V. Hardy and C. Mcloughlin.

**Figure captions:**

Figure 1. (Color online) Temperature dependence of magnetization measured at 50 Oe applied field along (a) In-plane and (b) Out-of-plane direction under ZFC [open circles] and FC [filled circles] conditions. Inset in (a) shows the first derivative of magnetization versus temperature where $T^*$ corresponds to the Néel temperature, and $T_{SRO}$ and $T_{LSMO}$ being, Curie temperatures of $SrRuO_3$ and $La_{0.7}Sr_{0.3}MnO_3$, respectively.

Figure 2. (Color online) (a): Magnetic hysteresis loops recorded at 70 K in in-plane (circles) and out-of-plane (squares) orientation [The low field regime is clearly shown in the inset], (b) and (c): Isothermal magnetization curves at different temperatures under (b) In-plane and (c) Out-of-plane configurations. Inset of Fig. (b) shows the variation of $H_T$ as a function of temperature.

Figure 3. (Color online) Temperature dependence of magnetic-entropy change ($-\Delta S_M$ ) under (a) In-plane and (b) Out-of-plane configurations, calculated from magnetization isotherms in the temperature range 50-165 K. The arrow mark shows that $\Delta S = 0$.

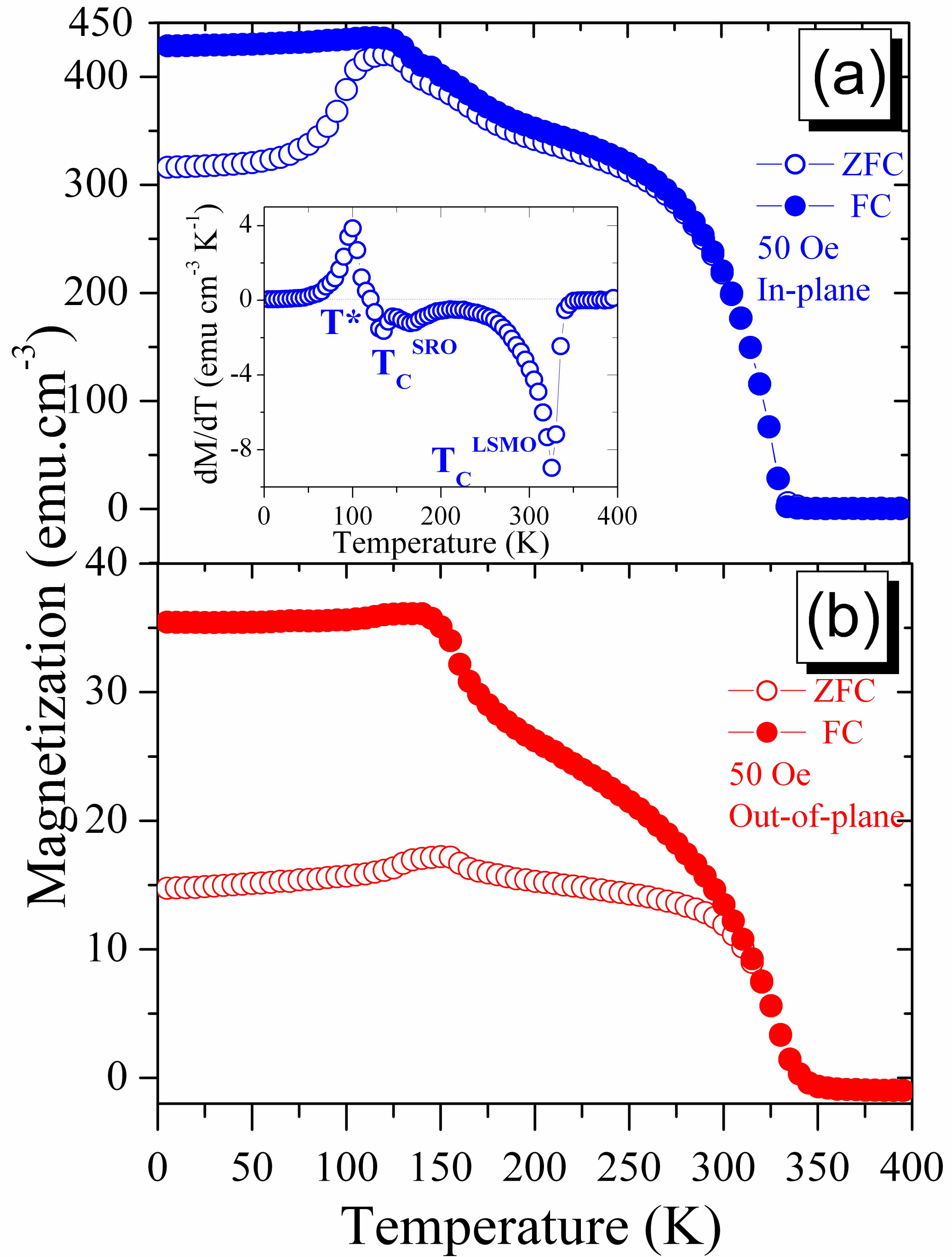

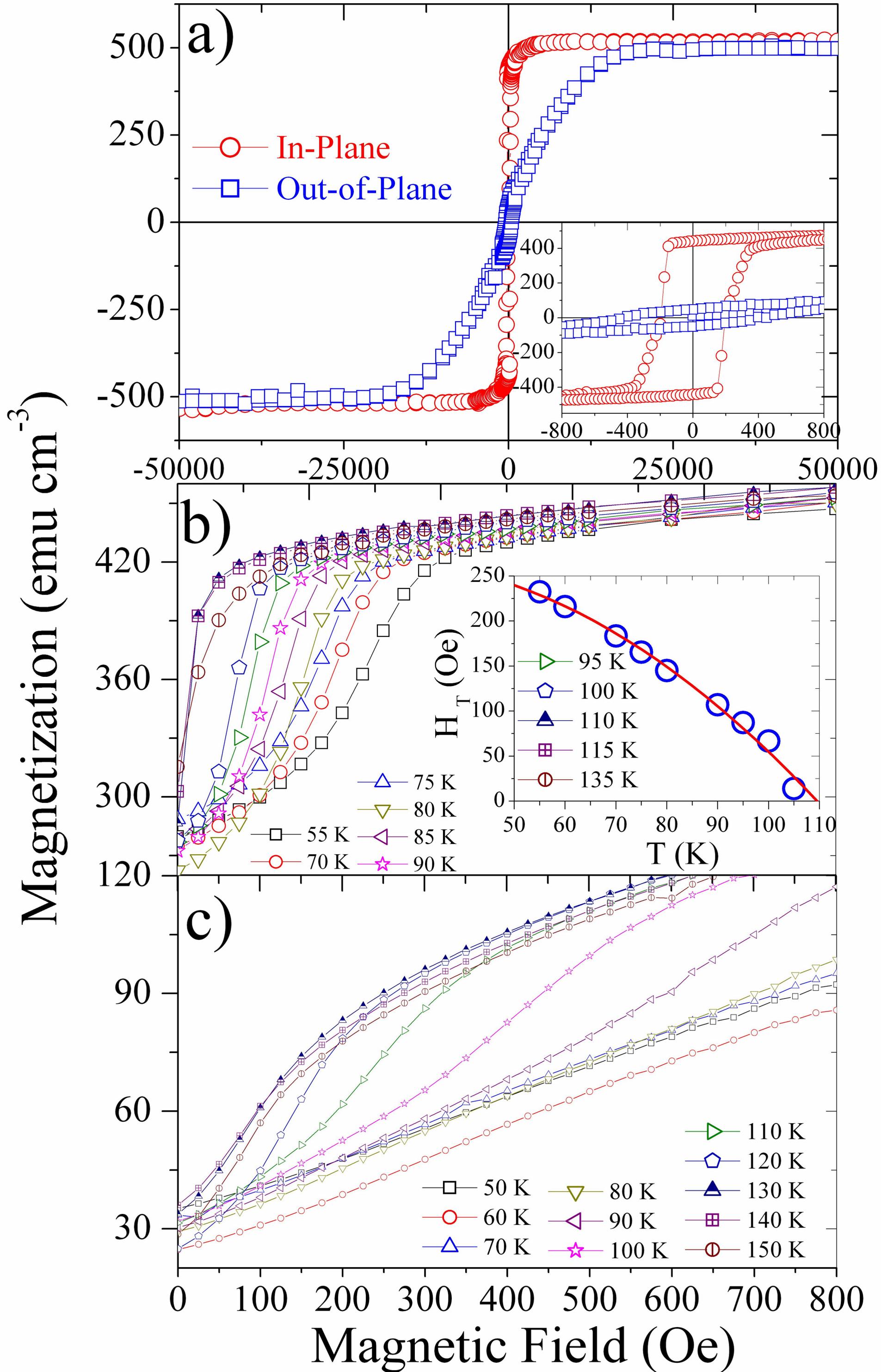

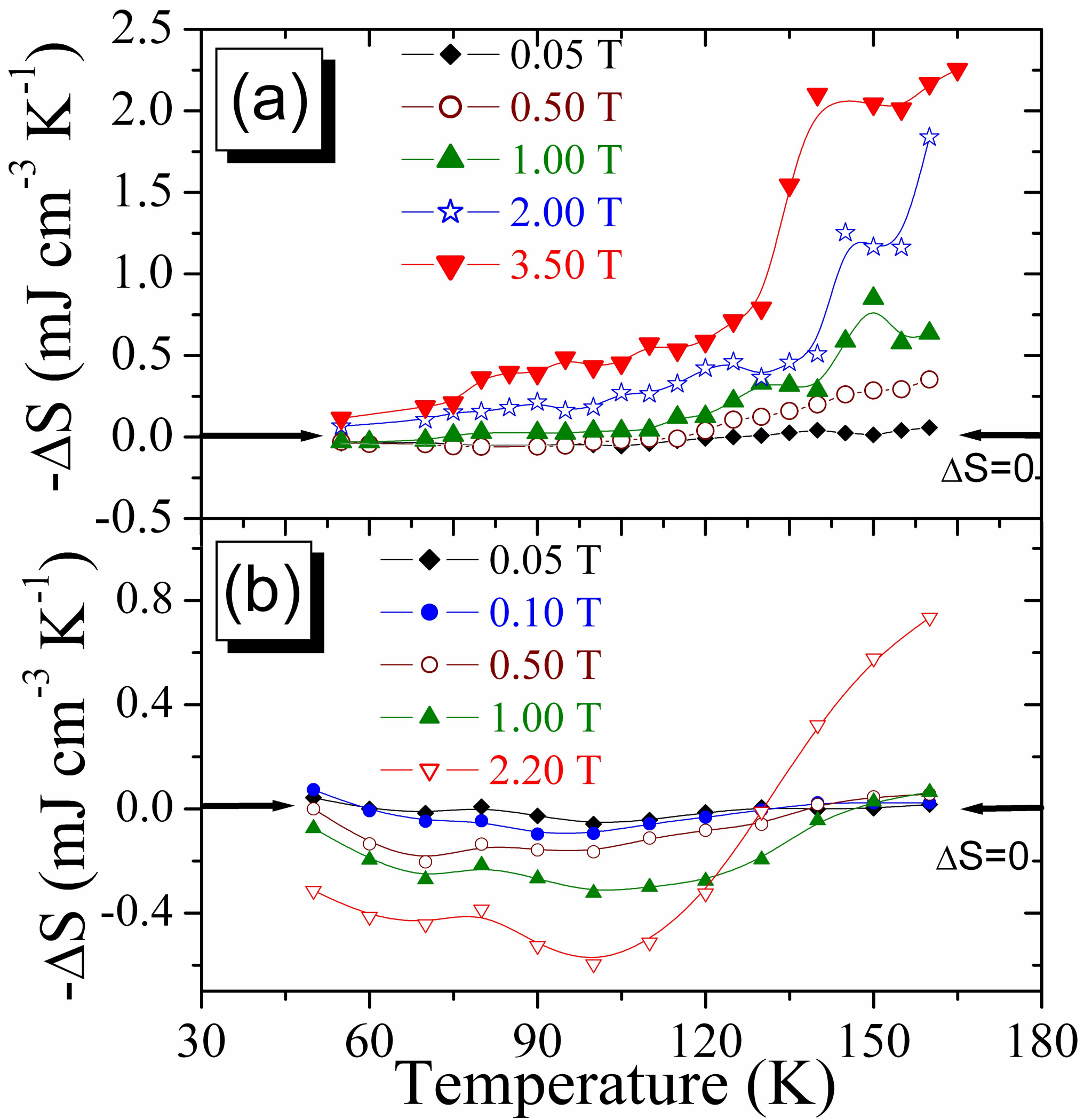